\documentstyle[aps,epsfig,pre]{revtex}
\begin{document}

\title{Nonequilibrium phase transitions induced by multiplicative
noise: effects of self-correlation}

\author{Sergio E. Mangioni\thanks{smangio@mdp.edu.ar} and Roberto R.
Deza\thanks{deza@mdp.edu.ar}}
\address{Departamento de F\'{\i}sica, Facultad de Ciencias Exactas y
Naturales \\ Universidad Nacional de Mar del Plata\\ De\'an Funes
3350, 7600 Mar del Plata, Argentina.}

\author{Ra\'ul Toral\thanks{dfsrtg0@ps.uib.es, WWW
http://formentor.uib.es/$\sim$raul}}
\address{Departament de F\'{\i}sica, Universitat de les Illes Balears
and\\ Instituto Mediterr\'aneo de Estudios Avanzados, IMEDEA
(CSIC-UIB)\\ E-07071 Palma de Mallorca, Spain}

\author{Horacio S. Wio\thanks{Member of CONICET, Argentina, ICTP
Associate, E-mail: wio@cab.cnea.gov.ar}}
\address{Centro At\'omico Bariloche, Comisi\'on Nacional de
Energ\'{\i}a At\'omica, and Instituto Balseiro\thanks{From Comisi\'on
Nacional de Energ\'{\i}a At\'omica and Universidad Nacional de Cuyo}\\
8400 San Carlos de Bariloche, Argentina.}
\date{\today}
\maketitle

\begin{abstract}
A recently introduced lattice model, describing an extended system
which exhibits a {\em reentrant\/} (symmetry-breaking, second-order)
noise-induced nonequilibrium phase transition, is studied under the
assumption that the multiplicative noise leading to the transition is
{\em colored}.  Within an effective Markovian approximation and a
mean-field scheme it is found that when the self-correlation time
$\tau$ of the noise is different from zero, the transition is {\em
also reentrant\/} with respect to the spatial coupling $D$.  In other
words, at variance with what one expects for equilibrium phase
transitions, a large enough value of $D$ favors {\em disorder}.
Moreover, except for a small region in the parameter subspace
determined by the noise intensity $\sigma$ and $D$, an increase in
$\tau$ usually {\em prevents\/} the formation of an ordered state.
These effects are supported by numerical simulations.
\end{abstract}
\pacs{PACS numbers: 47.20.Ky, 05.40.+j, 47.20.Hw}

\section{Introduction}
In the last few decades we have witnessed a paradigmatic shift
regarding the role of fluctuations, from the equilibrium picture of
merely being a $N^{-1/2}$ perturbation on thermodynamic averages---or
triggering at most phase transitions between well defined minima of
the free energy \cite{Pa}---to lead a host of new and amazing
phenomena in far from equilibrium situations.  As examples we may cite
noise-induced unimodal-bimodal transitions in some zero-dimensional
models \cite{HL}, stochastic resonance in zero-dimensional and
extended systems \cite{M,W}, noise-induced spatial patterns
\cite{ga93}, noise-delayed decay of unstable states \cite{ag98},
ratchets \cite{ratchets}, shifts in critical points \cite{shifts},
etc.

Recently it has been shown that a white and Gaussian {\em
multiplicative\/} noise can lead an {\em extended\/} dynamical system
(fulfilling appropriate conditions) to undergo a {\em phase\/}
transition towards an {\em ordered\/} state, characterized by a
nonzero order parameter and by the breakdown of ergodicity \cite{VPT}.
This result---first obtained within a Curie--Weiss-like mean-field
approximation, and further extended to consider the simplest
correlation function approach---has been confirmed through extensive
numerical simulations \cite{VPTK}.  In addition to its {\em
critical\/} nature as a function of the noise intensity $\sigma$, the
newly found noise-induced phase transition has the noteworthy feature
of being {\em reentrant}: for each value of $D$ above a threshold one,
the ordered state exists only inside a window $[\sigma_1,\sigma_2]$.
At variance with the known case of {\em equilibrium\/} order-disorder
transitions that are induced (in the simplest lattice models) by the
nearest-neighbor coupling constant $D$ and rely on the bistability of
the local potential, the transition in the case at hand is led by the
{\em combined effects\/} of $D$ and $\sigma$ through the
nonlinearities of the system.  Neither the zero-dimensional system
(corresponding to the $D=0$ limit) nor the deterministic one
($\sigma=0$) show any transition.

This counterintuitive ordering role of noise has also been found
afterwards in different models in the literature
\cite{ga96,VPAH,ge98,ki97,mu97,ga98}.  In Refs.\cite{ga96,VPAH}, the
authors study a noise-induced reentrant transition in a time-dependent
Ginzburg--Landau model with both additive and multiplicative noises.
Ref.\cite{ge98} introduces another simple model with a purely
multiplicative noise, which also presents a noise-induced reentrant
transition.  This reference also gives evidence that the universality
class of its critical behavior is that of the {\em multiplicative
noise} \cite{gr96,tu97} (see also Ref.\cite{ra95} for a discussion of
the universality class of these models).  In Ref.\cite{ki97}, a
first-order phase transition induced by noise is obtained in a system
of globally coupled oscillators.  A similar first-order phase
transition is also found in Ref.\cite{mu97}.  Finally, Ref.\cite{ga98}
addresses the role of multiplicative noise in the context of
phase-separation dynamics.

Although for the sake of mathematical simplicity all these references
(in particular, Refs.\cite{VPT,VPTK}) studied only the white-noise
case (the only exception is reference \cite{ga98} in which colored
noise in space, white in time, is considered) it is expected that,
because of their nature, fluctuations coupled {\em multiplicatively\/}
to the system will show some degree of time-correlation or ``color''
\cite{HL,S,SS,DL,HJ}, and hence new effects may arise from this fact.
For example, a reentrant behavior has been found recently as a
consequence of color in a noise-induced transition \cite{CSW} and an
ordering nonequilibrium phase transition can be induced in a
Ginzburg--Landau model by varying the correlation time of the additive
noise \cite{ga92,ga94}.  Thus motivated, we have studied the
consequences of a {\em finite\/} (but still very short as compared
with the ``deterministic'' or coarse-grained time scales)
self-correlation time $\tau$ of the multiplicative noise in systems of
this kind.  We now recall some of the new effects that emerge in this
colored-noise case, which have been briefly reported in a recent work
\cite{MDWT}:
\begin{itemize}
\item Our main finding is that, as a consequence of the multiplicative
character of the noise, a strong enough spatial coupling $D$ leads
invariably (for $\tau>0$) to a {\em disordered\/} state, contrary to
what would be expected to happen in equilibrium statistical-mechanical
models.

\item Another important result is that, except for large values of
$\sigma$ and {\em very small\/} values of $\tau$, color has an {\em
inhibiting\/} role for ordered states.  Moreover, there exists a {\em
finite\/} and not very large value of $\tau$ beyond which order is
impossible.
\end{itemize}
These results represent a major departure from what one can expect on
the basis of equilibrium thermodynamics, according to which one should
tend to think that as $D\to\infty$ an {\em ordered\/} situation is
favored.  Whereas that ``intuitive'' notion remains valid if the
multiplicative noise that induces the nonequilibrium ordering phase
transition is white \cite{VPT,VPTK}, it ceases to be so for $\tau>0$.
In the former case, the results could be interpreted in terms of a
``freezing'' of the short-time behavior by a strong enough spatial
coupling: for $D/\sigma^2\to\infty$, the stationary probability
distribution could be considered to be $\delta$-like, just as the
initial one.  In our case, an analysis of the short-time behavior of
the order parameter up to first order in $\tau$ reveals that the
disordering effect of $D$ can only be felt for {\em nonzero\/}
self-correlation time.  As $\tau$ increases, the minimum value of $D$
required to stabilize the disordered phase decreases and the region in
parameter space available to the ordered phase shrinks until it
vanishes.  Thus, the foregoing results can only be interpreted once we
recall that the ordered phase arises from the cooperation of {\em
two\/} factors: the presence of spatial coupling {\em and\/} the
multiplicative character of the noise (which may eventually lead to
``counterintuitive'' results).

It is our purpose in this work to render an explicit account of our
calculation and, at the same time, to expose and to discuss the
results more thoroughly.  After presenting the model in section
\ref{model}, we begin section \ref{approx} by introducing the
approximations needed to render the problem accessible to mathematical
analysis.  We resort to a mean-field approximation like in
Refs.\cite{VPT,VPTK} and to a ``unified colored noise approximation"
(UCNA) \cite{JH,CWA}, devised to deal with self-correlated noises.  In
section \ref{simpler} a simplified treatment using the aforementioned
approximations is given and in section \ref{refined} we expose the
more sophisticated approach that was actually used to obtain the phase
diagrams.  In this approach the UCNA is complemented with an
appropriate interpolation scheme \cite{CWA}.  In section \ref{result}
we expose and discuss the results obtained within the last approach,
comparing the phase diagram with the ones arising from the simplified
approach and (for $\tau\simeq 0$) from a perturbative expansion, and
the $D$-dependence of the order parameter $m$ for nonzero $\tau$ with
a numerical simulation \cite{MDWT}.  A final discussion of the
approach and its results is made in section \ref{conclu}.

\section{The model}\label{model}
The model under consideration has been introduced in
Refs.\cite{VPT,VPTK,MDWT}: a $d$-dimensional extended system of
typical linear size $L$ is restricted to a hypercubic lattice of
$N=L^d$ points, whereas time is still regarded as a continuous
variable.  The state of the system at time $t$ is given by the set of
stochastic variables $\{x_i(t)\}$ ($i=1,\dots,N$) defined at the sites
${\bf r}_i$ of this lattice, which obey a system of coupled ordinary
stochastic differential equations (SDE)
\begin{equation}\label{eq:1}
\dot{x}_i=f(x_i)+g(x_i)\eta_i+\frac{D}{2d}\sum_{j\in n(i)}(x_j-x_i)
\end{equation}
(throughout the paper, the Stratonovich interpretation for the SDE's
will be meant).  Eqs.(\ref{eq:1}) are the discrete version of the {\em
partial\/} SDE which in the continuum would determine the state of the
extended system: we recognize in the first two terms the
generalization of Langevin's equation for site $i$ to the case of
multiplicative noise ($\eta_i$ is the {\em colored\/} multiplicative
noise acting on site ${\bf r}_i$).  For the specific example analyzed
in Ref.\cite{VPT}, perhaps the simplest one exhibiting the transition
under analysis (see, however, Ref.\cite{ge98}),
\begin{equation}\label{eq:2}
f(x)=-x(1+x^2)^2,\hspace{2.0cm}g(x)=1+x^2.
\end{equation}
The last term in Eqs.(\ref{eq:1}) is nothing but the lattice version
of the Laplacian $\nabla^2 x$ of the extended stochastic variable
$x({\bf r},t)$ in a reaction-diffusion scheme.  $n(i)$ stands for the
set of $2d$ sites which form the immediate neighborhood of the site
${\bf r}_i$, and the coupling constant $D$ between neighboring lattice
sites is the diffusion coefficient.

As previously stated, it is our aim in this work to investigate the
effects of the self-correlation time $\tau$ of the multiplicative
noise on the model system just described.  To that end we must assume
a specific form for the noises $\{\eta_i\}$: we choose
Ornstein--Uhlenbeck noises, i.e.\ Gaussian distributed stochastic
variables with zero mean and exponentially decaying correlations:
\begin{equation}\label{eq:3}
\langle\eta_i(t)\eta_j(t')\rangle=\delta_{ij}(\sigma^2/2\tau)
\exp(-|t-t'|/\tau).
\end{equation}
They arise as solutions of an {\em uncoupled\/} set of Langevin SDE:
\begin{equation}\label{eq:4}
\tau\dot{\eta}_i=-\eta_i+\sigma\xi_i
\end{equation}
where the $\{\xi_i(t)\}$ are white noises---namely, Gaussian
stochastic variables with zero mean and $\delta$-correlated:
$\langle\xi_i(t)\xi_j(t')\rangle=\delta_{ij}\delta(t-t')$.  For
$\tau\to 0$, the Ornstein--Uhlenbeck noise $\eta_i(t)$ approaches the
white-noise limit $\xi^W_i(t)$ with correlations $\langle\xi^W_i(t)
\xi^W_j(t')\rangle=\sigma^2\,\delta_{ij}\,\delta(t-t')$, which was the
case studied in Refs.\cite{VPT,VPTK}.

\section{The approximations}\label{approx}
The non-Markovian character of the process $\{x_i\}$ (due to the
presence of the colored noise $\{\eta_i\}$) makes it difficult to
study.  However, some approximations have been devised which render a
{\em Markovian\/} process that---whereas nicely simplifying the
treatment---can still capture some of the essential features of the
complete non-Markovian one.  The aforementioned UCNA is one of them:
in fact a very reliable one, because of its ability to reproduce the
small-- and large-$\tau$ limits \cite{JH}.  By resorting to
interpolation schemes, one can retrieve meaningful results in wider
vicinities of these limits \cite{CWA}.

As already declared, our approach is a mean-field-like one.  The
earlier we make this kind of assumptions in the calculation, the
cruder the approximation will turn out to be.  In order to find the
phase diagram in the presence of colored noise we have made the
mean-field approximation at some late stage, so enhancing the
precision of the calculation.  However, since this calculation is a
tedious one, we shall first expose a simpler approximation which
brings out most qualitative results.  We aim in this way to underline
the physical origin of the results presented in section IV. The
differences arising from both calculations are pointed out there.

\subsection{A simpler approach}\label{simpler}
The simpler mean-field approximation follows closely Curie--Weiss'
mean-field approach to magnetism, and consists in replacing the last
term in Eqs.(\ref{eq:1})
\begin{equation}\label{eq:5}
\Delta_{i}\equiv\frac{D}{2d}\sum_{j\in n(i)}(x_j-x_i)
\end{equation}
by
\begin{equation}\label{eq:6}
\bar\Delta_{i}\equiv D(\bar{x}-x_i),
\end{equation}
where $\bar{x}$ is a {\em parameter\/} that will be determined
self-consistently.  In other words, the (short-ranged) interactions
are substituted by a time-- and space-independent ``external'' field
whose value {\em depends on the state\/} of the system.  Since in this
approximation Eqs.(\ref{eq:1}) get immediately decoupled, there is no
use in keeping the subindex $i$ and we may refer to the systems in
Eqs.\ (\ref{eq:1}) and (\ref{eq:4}) as if they were single equations.
Hereafter, the primes will indicate derivatives with respect to $x$
(clearly $\bar\Delta'=-D$).

If we take the time derivative of Eq.(\ref{eq:1}), replace first
$\dot{\eta}$ in terms of $\eta$ and $\xi$ from Eq.(\ref{eq:4}) and
then $\eta$ in terms of $\dot x$ and $x$ from Eq.(\ref{eq:1}), we
obtain the following {\em non-Markovian\/} SDE:
\begin{equation}\label{eq:7}
\tau(\ddot{x}-\frac{g'}{g}
\dot{x}^2)=-\left(1-\tau\left[(f+\bar\Delta)'-\frac{g'}{g}
(f+\bar\Delta)\right]\right)\dot{x}+(f+\bar\Delta)+\sigma g\xi.
\end{equation}
The aforementioned UCNA allows us to recover a Markovian SDE: for our
particular problem it amounts, on one hand, to a usual adiabatic
elimination (namely, neglecting $\ddot x$) and, on the other, to
neglect ${\dot x}^2$ so that the system's dynamics be governed by a
Fokker--Planck equation \cite{WCSPR}.  The resulting equation, being
{\em linear\/} in $\dot x$ (but of course not in $x$), can be
immediately solved for $\dot x$ giving
\begin{equation}\label{eq:8}
\dot x=Q(x;\bar{x})+S(x;\bar{x})\xi,
\end{equation}
with
\begin{eqnarray}
Q(x;\bar{x})&\equiv&(f+\bar\Delta)\theta,\label{eq:9}\\
S(x;\bar{x})&\equiv&\sigma g\theta,\label{eq:10}\\
\theta(x;\bar{x})&\equiv&\{1-\tau g[(f+\bar\Delta)/g]'\}^{-1}
\label{eq:11}.
\end{eqnarray}
The parametric dependence of $Q(x)$ and $S(x)$ on $\bar{x}$ has been
written explicitly.

The Fokker--Planck equation associated to the SDE Eq.(\ref{eq:8}) is
\begin{equation}
\label{eq:12}
\partial_t P(x,t;\bar{x})=
-\partial_x\left[R_1(x;\bar{x})P(x,t;\bar{x})\right]+
\frac{1}{2}\partial^2_x\left[R_2(x;\bar{x})P(x,t;\bar{x})\right],
\end{equation}
with drift and diffusion coefficients given by \cite{risken}:
\begin{eqnarray}
\label{eq:13}
R_1(x;\bar{x})&=&Q+\frac{1}{4}(S^2)'\\
\label{eq:14}
R_2(x;\bar{x})&=&S^2.
\end{eqnarray}
The solution of the time-independent Fokker--Planck equation leads to
the stationary probability density
\begin{equation}\label{eq:15}
P^{st}(x;\bar{x})={\cal N}^{-1}\exp{\left[\int_{0}^{x}dx'
\frac{2R_1(x';\bar{x})-\partial_{x'}R_2(x';\bar{x})}{R_2(x';\bar{x})}
\right]},
\end{equation}
being ${\cal N}$ its norm.  The partial-derivative notation
$\partial_{x'}$ in Eq.(\ref{eq:15})---as in Eqs.\ (\ref{eq:18}) and
(\ref{eq:19}) below---is only a reminder of the parametric dependence
of $R_1$, $R_2$ on $\bar{x}$.

The value of $\bar{x}$ arises from a self-consistency relation, once
we equate it to the average value of the random variable $x_i$ in the
stationary state
\begin{equation}\label{eq:16}
\bar{x}=\langle x\rangle\equiv\int_{-\infty}^{\infty}dx\,x\,
P^{st}(x;\bar{x})\equiv F(\bar{x}).
\end{equation}
As in the known Curie--Weiss mean-field approach, the condition
\begin{equation}\label{eq:17}
\left.\frac{dF}{d\bar{x}}\right|_{\bar{x}=0}=1
\end{equation}
allows us to find the transition line between the ordered and the
disordered phases.  Here also, $F(\bar{x})$ is a smooth odd function
such that Eq.(\ref{eq:16}) has always a root at $\bar{x}=0$ and for
$\left.dF/d\bar{x}\right|_{\bar{x}=0}>1$ it has two nontrivial roots
which differ only in sign.  The condition Eq.(\ref{eq:17}) thus reads:
\begin{equation}\label{eq:18}
{\cal N}^{-1}\int_{-\infty}^{\infty}dx\:x\int_0^x dx'\exp
{\left[\int_{0}^{x'}dx''\:\frac{2R_1-\partial_{x''}R_2}{R_2}\right]}
\partial_{\bar{x}}\left.\left(\frac{2R_1-\partial_{x'}
R_2}{R_2}\right)\right|_{\bar{x}=0}=1,
\end{equation}
where
\begin{equation}\label{eq:19}
\left.{\cal N}=\int_{-\infty}^{\infty}dx\exp{\left[\int_{0}^{x}dx'
\frac{2R_1-\partial_{x'}R_2}{R_2}\right]}\right|_{\bar{x}=0}.
\end{equation}
Eqs.\ (\ref{eq:18}) and (\ref{eq:19}) must be solved numerically in
order to find the lines in parameter space ($\sigma$,$\tau$,$D$) that
separate ordered ($\bar{x}\neq 0$) from disordered ($\bar{x}=0$)
phases.  The results of this calculation will be shown in section V.
Next, we introduce a more refined approach in which the mean-field
approximation is made at a later stage in the calculation.

\subsection{A more refined approach}\label{refined}
As we shall see, the relations obtained with this more sophisticated
approach are similar to Eqs.\ (\ref{eq:15}) through (\ref{eq:19}), but
with different expressions for the functions $R_1(x;\bar{x})$ and
$R_2(x;\bar{x})$.  The idea here is to {\em introduce first\/} the
UCNA, without yet resorting to the mean-field approximation.  In the
following, $\Delta_i$ has the same meaning as in Eq.(\ref{eq:5}) and,
as it occurred previously with $\bar\Delta'$, it satisfies
$\Delta_i'=-D$.  Note however that whereas for $f_i\equiv f(x_i)$ and
$g_i\equiv g(x_i)$ the prime has the meaning of a total derivative
with respect to $x_i$, for $\Delta_i'$ and all the functions involving
them, its meaning is really that of a {\em partial\/} derivative with
respect to $x_i$.  Proceeding as before, i.e.\ taking the time
derivative of Eqs.(\ref{eq:1}) and using Eqs.\ (\ref{eq:1}) and
(\ref{eq:4}) to eliminate the $\eta$'s in favor of the $x$'s and
$\xi$'s, we obtain the following system of (non Markovian) SDE's
\begin{equation}\label{eq:20}
\tau(\ddot{x}_i-\frac{g'_i}{g_i}\dot{x}_i^2)=-\left[1-\tau
g_i\left(\frac{f_i+\Delta_i}{g_i}\right)'\right]{\dot
x}_i+(f_i+\Delta_i)+\sigma g_i\xi_i+\frac{D\tau}{2d}\sum_{j\in
n(i)}{\dot x}_j.
\end{equation}

The UCNA proceeds here through the neglect of ${\ddot x}_i$ and of
$({\dot x}_i)^2$, so retrieving a {\em linear\/} equation in the $\dot
x$'s (but of course not in the $x$'s), which can be rewritten as
\begin{eqnarray}\label{eq:21}
{\dot x}_i&=&
\left[1-\tau g_i\left(\frac{f_i+\Delta_i}{g_i}\right)'\right]^{-1}
\left[(f_i+\Delta_i)+\sigma g_i\xi_i+\frac{D\tau}{2d}
\sum_{j\in n(i)}{\dot x}_j\right]\nonumber\\
&=&[Q_i+S_i\xi_i]+\frac{\beta_i}{2d}\sum_{j\in n(i)}{\dot x}_j.
\end{eqnarray}
Here the quantities
\begin{eqnarray}
\theta_i&\equiv&\left[1-\tau
g_i\left(f_i+\Delta_i/g_i\right)'\right]^{-1},\label{eq:22}\\
Q_i&\equiv & (f_i+\Delta_i) \theta_i,\label{eq:23}\\
S_i&\equiv & \sigma g_i\theta_i,\label{eq:24}\\
\beta_i&\equiv&D\tau\theta_i\label{eq:25}
\end{eqnarray}
have been defined in order to simplify the notation.  Note that
although only the dependence upon $x_i$ has been made explicit in this
notation, these quantities depend also (through $\Delta_i$) on the
values $x_j$ at the neighboring sites.

Now, assuming the lattice to be isotropic, we apply to this set a {\em
mean-field-like approximation\/} (but not yet the main one) consisting
in {\em replacing\/} in all the functions appearing in
Eq.(\ref{eq:21}) the $2d$ neighbors $x_j$ of the variable $x_i$ {\em
by a common value\/} $y_i$.  Through this procedure one reduces to
{\em two\/} the number of different coupled SDE's: one for $x\equiv
x_i$ and another for its nearest neighbor variable $y\equiv y_i$.
These are
\begin{equation}
{\dot a}=h_a+g_{ab}\,\xi_b,\label{eq:26}
\end{equation}
where a sum over the values $x,y$ is implied for the indices $a,b$,
and the noise variables satisfy $\langle\xi_a(t)\xi_b(t')\rangle=
\delta_{ab}\,\delta(t-t')$.  If, similarly as before, we define
\begin{eqnarray}
\theta(x,y)&=&\left[1-\tau g(x)\frac{\partial}{\partial x}
\left(\frac{f(x)+D(y-x)}{g(x)}\right)\right]^{-1},\label{eq:27}\\
Q(x,y)&=&[f(x)+D(y-x)]\theta(x,y),\label{eq:28}\\
S(x,y)&=&\sigma g(x)\theta(x,y),\label{eq:29}\\
\beta(x,y)&=&\tau D\theta(x,y),\label{eq:30}\\
A(x,y)&=&[1-\beta(x,y)\beta(y,x)]^{-1}=A(y,x),\label{eq:31}
\end{eqnarray}
and write ${\bar a}=y$ if $a=x$ and viceversa, then the explicit forms
of the functions in Eq.(\ref{eq:26}) are
\begin{equation}\label{eq:32}
h_a=A(x,y)\,[Q(a,{\bar a})+\beta(a,{\bar a})Q({\bar a},a)]
\end{equation}
and
\begin{eqnarray}
g_{ab}&=&A(x,y)\,S(a,{\bar a})\qquad\qquad\mbox{ if }b=a,
\label{eq:34}\\
&=&A(x,y)\,\beta(a,{\bar a})S({\bar a},a)\quad\mbox{ if }b={\bar a}.
\label{eq:36}
\end{eqnarray}
The bivariate Fokker--Planck equation associated to Eqs.(\ref{eq:26})
is
\begin{equation}\label{eq:38}
\partial_t P=-\partial_a(R_aP)
+\frac{1}{2}\partial_a\partial_b(R_{ab}P),
\end{equation}
where $P=P(x,y;t)$.  According to standard techniques \cite{risken},
the drift and diffusion coefficients are given respectively by
\begin{eqnarray}
R_a(x,y)&=&h_a+\frac{1}{2}g_{bc}\partial_b(g_{ac}),\label{eq:39}\\
R_{ab}(x,y)&=&g_{ac}g_{bc}.\label{eq:40}
\end{eqnarray}
Since the denominators occurring in these equations may become zero
for some values of $x$ or $y$, we resort to an {\em interpolation
procedure\/} (analogous to the one used in Refs.\cite{CSW,CWA})
consisting in replacing the expression (\ref{eq:31}) for $A(x,y)$ by
\begin{equation}\label{eq:41}
A(x,y)=\frac{1-\beta(x,y)\beta(y,x)}{1+\beta(x,y)^2\beta(y,x)^2}.
\end{equation}
Since $\beta(x,y)$ is proportional to $\tau$, it follows that the
expression in Eq.(\ref{eq:31}) coincides with the interpolated one,
Eq.(\ref{eq:41}), as $\tau\to 0$ and $\tau\to\infty$ (the latter limit
meaning indeed ``$\tau$ comparable with the `deterministic' time
scales'').

By integrating the bivariate Fokker--Planck Eq.(\ref{eq:38}) with
respect to $y$ we obtain a single-variable equation, which in the
stationary case reads
\begin{equation}\label{eq:42}
0=-\partial_x(R_1 P^{st})+\frac{1}{2}\partial^2_x(R_2 P^{st}),
\end{equation}
being $R_1 P^{st}$ and $R_2 P^{st}$ functions of $x$ only:
\begin{eqnarray}
R_1 P^{st}&\equiv&\int_{-\infty}^{\infty}dy\,P^{st}(x,y)\,
R_x(x,y),\label{eq:43}\\
R_2 P^{st}&\equiv&\int_{-\infty}^{\infty}dy\,P^{st}(x,y)\,
R_{xx}(x,y).\label{eq:44}
\end{eqnarray}

Here it is when we resort to the {\em main mean-field-type
approximation}, resembling the Curie--Weiss' type of approach used
in Ref.\cite{VPT}: {\em assuming\/} $P^{st}(x,y)\approx
P^{st}(x)\delta(y-\bar{x})$, $R_1$ and $R_2$ in Eqs.\
(\ref{eq:43}) and (\ref{eq:44}) are approximated by
\begin{eqnarray}
R_1&=&R_x(x;\bar{x}),\label{eq:45}\\
R_2&=&R_{xx}(x;\bar{x}).\label{eq:46}
\end{eqnarray}
In this way, from the stationary {\em joint\/} probability density
function $P^{st}(x,y)$ we retrieve an {\em effective\/}
single-variable one $P^{st}(x;\bar{x})$ whose expression in terms of
$R_1(x;\bar{x})$ and $R_2(x;\bar{x})$ arising from Eq.(\ref{eq:42}) is
the same as in Eq.(\ref{eq:15}).  The value of $\bar{x}$ follows again
from a self-consistency relation like Eq.(\ref{eq:16}).  The procedure
to find the phase diagram is the same as in the foregoing subsection
and the explicit expression of the condition $\left.dF
/d\bar{x}\right|_{\bar{x}=0}=1$ is given by Eqs.\ (\ref{eq:18}) and
(\ref{eq:19}), this time in terms of the corresponding $R_1$, $R_2$
given by Eqs.\ (\ref{eq:45}) and (\ref{eq:46}).

As discussed in Ref.\cite{MDWT}, although the kind of approximation
leading to Eq.(\ref{eq:41}) is not of a perturbative nature, it has
provided sound results in the cases analyzed heretofore
\cite{CSW,CWA}.  Nonetheless, for the sake of comparison we have also
adapted to the case of a multiplicative noise a known perturbative
procedure \cite{ST}.  Within this context, the expressions for
$R_1(x;\bar{x})$ and $R_2(x;\bar{x})$ come from a consistent
small-$\tau$ expansion of a Fokker--Planck equation (Eqs.\
(\ref{eq:12}) or (\ref{eq:38})):
\begin{eqnarray}
R_1(x;\bar{x})&=&(f+\bar\Delta)+\sigma^2 x
\{g[1+\tau(f+\bar\Delta)']+\tau(f+\bar\Delta)\},\label{eq:47}\\
R_2(x;\bar{x})&=&\{\sigma g[1+\tau(f+\bar\Delta)']\}^2.\label{eq:48}
\end{eqnarray}
\section{The results}\label{result}
\subsection{Phase diagram}
In the following we shall describe the results obtained through the
numerical solution of Eqs.\ (\ref{eq:18}) and (\ref{eq:19}) in the
more refined approach, i.e.\ with $R_1$ and $R_2$ as prescribed by
Eqs.\ (\ref{eq:45}) and (\ref{eq:46}).  We shall also compare
these results with the ones arising from Eqs.\ (\ref{eq:13}) and
(\ref{eq:14}), and with a perturbative expansion for small $\tau$.
Figures \ref{mdtw1} to \ref{mdtw3} are respectively the projections
onto the $\sigma$--$D$, $\tau$--$\sigma$ and $\tau$--$D$ planes, of
cuts of the boundary separating the ordered and disordered phases
performed at fixed values of the remaining parameters.

Let us first concentrate on Fig.\ref{mdtw1}: it corresponds to Fig.1
in Ref.\cite{MDWT}, but is the result of an improved calculation based
on the more refined mean-field approach described in
Sect.\ref{refined}.  The novelty is that, at least for $\tau$ not too
small, it is now evident that the region available to the ordered
phase is {\em bounded\/}.  The noteworthy aspects are the following:
\renewcommand{\theenumi}{\Alph{enumi}}
\begin{enumerate}
\item\label{a} As in the white-noise case $\tau=0$
(Refs.\cite{VPT,VPTK}), the ordering phase transition is {\em reentrant
with respect to $\sigma$}: for a range of values of $D$ that depends
on $\tau$, ordered states can only exist within a window
$[\sigma_1,\sigma_2]$.  The fact that this window shifts to the right
{\em for small $\tau$\/} means that, for fixed $D$, color {\em
destroys\/} order just above $\sigma_1$ but {\em creates\/} it just
above $\sigma_2$.

\item\label{b} For fixed $\sigma>1$ and $\tau\ne 0$, ordered states
exist {\em only within a window\/} of values for $D$.  Thus the
ordering phase transition is {\em also reentrant with respect to $D$}.
For $\tau$ small enough the maximum value of $D$ compatible with the
ordered phase increases rather steeply with $\sigma$, reaching a
maximum around $\sigma\sim 5$ and then decreases gently.  For
$\tau\geq 0.1$ it becomes evident (in the ranges of $D$ and $\sigma$
analyzed) that the region sustaining the ordered phase is {\em
closed\/}, and shrinks to a point for a value slightly larger than
$\tau=0.123$.

\item\label{c} For fixed values of $\sigma>1$ and $D$ larger than its
minimum for $\tau=0$, the system {\em always\/} becomes disordered for
$\tau$ large enough.  The maximum value of $\tau$ consistent with
order altogether corresponds to $\sigma\sim 5$ and $D\sim 32$.  In
other words, ordering is possible {\em only\/} if the multiplicative
noise inducing it has short memory.

\item\label{d} The fact that the region sustaining the ordered phase
finally shrinks to a point means that even for that small region in
the $\sigma$--$D$ plane for which order is induced by color, a further
increase in $\tau$ destroys it.  In other words, the phase transition
is {\em also reentrant with respect to $\tau$}.  For $D$ large enough
there may exist even {\em two\/} such windows.
\end{enumerate}

Some of the features just described become more evident by looking at
Fig.\ref{mdtw2}:
\begin{itemize}
	\item the existence of a maximum correlation time consistent with
	ordering for each value of $D$ (occurring for an optimal value of
	$\sigma$) (\ref{c});

	\item the ordering ability (\ref{a}) of a very small amount of
	color for $\sigma>\sigma_2(D)$ ($\sigma_2(\tau,D)$ increases very
	rapidly at first);

	\item the reentrance with respect to $\tau$ and even the
	occurrence of a {\em double\/} reentrance for $D$ large enough
	(\ref{d}).
\end{itemize}
Fig.\ref{mdtw3} represents another way of seeing the reentrance with
respect to $D$ for constant $\sigma$ (large enough) and the fact that
there exists a maximum $\tau$ consistent with order for each value of
$\sigma$ (being it largest for $\sigma\sim 5$).  The scarce dependence
of $D$ on $\tau$ in the lower branch---as well as its almost linear
dependence on $\sigma$---is easily understood by looking at the
rightmost branch of Fig.\ref{mdtw1}.

In Figs.\ \ref{mdtw4} to \ref{mdtw6} we compare the results just
shown---obtained as we said using the the more refined approach of
Sect.\ref{refined}---with the ones arising from the simpler one
(Sect.\ref{simpler}).  Figure \ref{mdtw4} corresponds to
Fig.\ref{mdtw2}, whereas Figs.\ \ref{mdtw5} and \ref{mdtw6} focus
respectively on the $\tau=0.03$ and $\tau=0.015$ curves in
Fig.\ref{mdtw1}.  Not only does the simpler approach (grossly)
overestimate the size of the ordered region but also---as one may
infer from Figs.\ \ref{mdtw4} and \ref{mdtw5}---it seems to predict
unbounded ordered regions.

Figure \ref{mdtw6} corresponds to a rather small value of $\tau$, so
that a comparison with the results obtained by using expressions
(\ref{eq:47}) and (\ref{eq:48}) makes sense.  For $\sigma$ and $D$
small enough the curves almost coincide.  As it is well known, the
simultaneous consideration of small $\tau$ and large $\sigma$ cannot
be done independently \cite{SS,DL,HJ}.  In the present case, a similar
effect arises when we consider large values of $D$, as we discuss
below.  Hence, in order to consider larger values of $\sigma$ and $D$,
one should take {\em extremely low\/} values of $\tau$.  As
Fig.\ref{mdtw6} shows, already for $\tau=0.015$ there is an apparent
discrepancy between the perturbative results and the mean-field ones
even for not so large values of $\sigma$ and $D$.  The noteworthy fact
is that the perturbative expansion also tends to indicate the
existence of a reentrance with respect to $D$.

\subsection{Order parameter}
The order parameter in this system is $m\equiv|\bar{x}|$ namely, the
positive solution of the consistency equation (see Eq.(\ref{eq:16}) in
Sect.\ref{simpler}).  In Fig.\ref{mdtw7} we plot $m$ vs.\ $\sigma$ for
$D=20$ and different values of $\tau$.  Consistently with what has
been discussed in (\ref{a}) and (\ref{c}) and shown in
Fig.\ref{mdtw1}, we see that as $\tau$ increases the window of
$\sigma$ values where ordering occurs shrinks until it disappears.
One also notices that at least for this $D$, the value of $\sigma$
corrresponding to the maximum order parameter varies very little with
$\tau$.  Figure \ref{mdtw8} is a plot of $m$ vs.\ $\tau$ for $D=45$
and two values of $\sigma$ ($\approx 7.07$ and $\approx 8.94$) that
illustrates the case of double reentrance in $\tau$.

Since the previous results have been obtained in the mean-field
approximation, we have also performed numerical simulations in order
to have an independent check of the predictions.  As a representative
example---corresponding to the phenomenon (\ref{b}) above (the
destruction of the ordered phase by an increasing coupling constant
$D$)---we plot jointly in Fig.\ref{mdtw9}, for rather small values of
$\sigma$ ($=2$) and $\tau$ ($=0.01$), the $D$-dependence of the order
parameter as predicted by our mean-field approach and as resulting
from a numerical integration of the original SDE's, Eqs.(\ref{eq:1}).
We have taken three different lattice sizes in order to assess
finite-size effects.  As we see, the numerical simulations {\em do\/}
predict the disordering for large enough $D$, and even the maximum
ordering occurs in a region which is consistent with the mean-field
prediction.  This comparison warns us, however, that the mean-field
approximation can severely underestimate the size of the ordered
region.

The {\em short-time evolution\/} of $\langle x\rangle$ can be obtained
multiplying Eq.(\ref{eq:12}) by $x$ and integrating:
\begin{equation}\label{eq:49}
\frac{d\langle x\rangle}{dt}=\int_{-\infty}^{\infty} dx
R_1(x;\bar{x})P(x,t;\bar{x}).
\end{equation}
Let us assume an initial condition such that at early times $P(x,t\sim
0;\bar{x})=\delta(x-\bar{x})$.  Equating $\bar{x}=\langle x\rangle$ as
before, we obtain
\begin{equation}\label{eq:50}
\frac{d\langle x\rangle}{dt}=R_1(\bar{x},\bar{x})
\end{equation}
(again, we can use for $R_1$ the result Eq.(\ref{eq:13}) of the simple
approximation or the more elaborate one given by Eq.(\ref{eq:45})).
The numerical solution of Eq.(\ref{eq:50}) has an initial {\em
rising\/} period (it is initially {\em unstable\/}) reaching very soon
a maximum and tending to zero afterwards.

For $D/\sigma^2\to\infty$, Eq.(\ref{eq:50}) results to be valid also
in the {\em asymptotic regime\/} since $P^{st}(x)=\delta(x-\bar{x})$
\cite{VPTK}.  In Ref.\cite{VPTK} an equivalent equation is obtained in
the $\tau=0$ case for {\em both\/} limits ($D=0$ {\em and\/}
$D/\sigma^2\to\infty$) being there interpreted in terms of a
``freezing'' of the short-time behavior.  According to this criterion,
in the $D/\sigma^2\to\infty$ limit the system undergoes a second-order
phase transition {\em if\/} the corresponding zero-dimensional model
presents {\em a linear instability in its short-time dynamics\/},
i.e.\ if after linearizing Eq.(\ref{eq:50}):
\begin{equation}\label{eq:51}
\langle\dot x\rangle=-\alpha\langle x\rangle
\end{equation}
one finds that $\alpha<0$.  We then see that the trivial (disordered)
solution $\langle x\rangle=0$ is stable only for $\alpha>0$.  For
$\alpha<0$ other stable solutions with $\langle x\rangle\ne 0$ appear,
and the system develops order through a genuine {\em phase\/}
transition.  In this case, $\langle x\rangle$ can be regarded as the
{\em order parameter}.  In the white noise limit $\tau=0$ this is
known to be the case for sufficiently large values of the coupling $D$
and for a window of values for the noise amplitude
$\sigma\in[\sigma_1,\sigma_2]$.

We discuss now how the stability of the ordered phase is altered by
nonzero values of $\tau$.  If we linearize Eq.(\ref{eq:50}) using the
expression of $R_1(\bar{x},\bar{x})$ from Eq.(\ref{eq:13}), we
obtain
\begin{equation}\label{eq:52}
\alpha=
\frac{(1+\tau+\tau D)^2-\sigma^2(1-3\tau+2\tau D)}{(1+\tau+\tau D)^3}.
\end{equation}
If we use instead Eq.(\ref{eq:45}), the result can be written
exclusively in terms of $\tau D$:
\begin{equation}\label{eq:53}
\alpha=1-\sigma^2\frac{B(\tau D)}{A(\tau D)}
\end{equation}
with
\begin{eqnarray}
	A(x)&=&1+5x+8x^2+3x^3-3x^4-x^5+2x^6+x^7,\label{eq:54}\\
	B(x)&=&1+3x+x^2-5x^3-3x^4+3x^5+x^6,\label{eq:55}
\end{eqnarray}
so the instability occurs at $\sigma^2=A/B$.  Now, the ratio $B/A$
keeps always below 1 in the positive range, has a minimum of $\sim
0.05$ at $\tau D\sim 1.09$ and a maximum of $\sim 0.36$ at $\tau D\sim
2.33$.  If we consider e.g.\ the conditions in Fig.\ref{mdtw5} (namely
$\tau=0.03$) and take $D$ large enough so that $\tau D>2.33$, one can
see that Eq.(\ref{eq:53}) approximates the left boundary better than
Eq.(\ref{eq:52}) does.  In the limit $\tau\ll 1$ (but still finite)
Eq.(\ref{eq:53}) can be approximated to the expression reported in
Ref.\cite{MDWT}, namely
\begin{equation}\label{eq:56}
\alpha=\frac{1+\tau D-\sigma^2}{1+\tau D}.
\end{equation}
It is worthwile to stress the fact (evident from Eqs.\ (\ref{eq:52}),
(\ref{eq:53}) and (\ref{eq:56})) that the value of $D$ has no effect
on the {\em location\/} of the instability by itself, but only through
the combination $\tau D$: according to Eq.(\ref{eq:56}) the stable
phase is the disordered one ($\langle x\rangle=0$) for $1+\tau
D>\sigma^2$ (since $\alpha>0$) and the ordered one ($\langle
x\rangle\neq 0$) for $1+\tau D<\sigma^2$.  In summary, whereas the
noise intensity $\sigma$ has a destabilizing effect on the disordered
phase, as soon as $\tau\neq 0$ the spatial coupling $D$ tends to {\em
stabilize\/} it.  For $\tau=0$ the last effect is not present, being
then $\sigma>1$ the condition for ordering \cite{VPT,VPTK}.
Considering that the effect of even a tiny correlation is enhanced by
$D$, we can understand the abrupt change in slope (from negative to
positive) shown in Fig.\ref{mdtw1} as soon as $\tau\neq 0$.  Note the
approximately inverse relation between $\tau$ and $D$ for fixed
$\sigma$ on the upper branches of Fig.\ref{mdtw3}, even when
Eq.(\ref{eq:51}) is strictly valid for $D/\sigma^2\to\infty$.

\section{Conclusions}\label{conclu}
This work has focused on the effects of a self-correlation in the
multiplicative noise, on the reentrant noise-induced phase transition
introduced in Ref.\cite{VPT}.  Whereas in a recent Letter we have
reported the most relevant results \cite{MDWT}, it has been our
purpose in the present work to expose in more detail the techniques
and the approximations employed.  We also discuss more thoroughly the
results, adding new figures and enriching the contents of others.

Through the use of the UCNA we recovered a Markovian behavior for the
system, and through an interpolation scheme similar to the one
introduced in Ref.\cite{CWA} we resolved indeterminacies in the
equations describing it.  We stress that the equation resulting from
this interpolation scheme are {\em exact\/} in the limits $\tau=0$ and
$\tau\to\infty$.  In addition to the fact that the interpolation
scheme has been already applied with success in other works
\cite{CSW,CWA}, the goodness of our approximations for small but
nonetheless finite values of $\tau$ has been checked against a
standard perturbative expansion \cite{ST} (adapted for multiplicative
noise).  It is worth to emphasize the fact that these approximations
are, so far, the {\em only\/} tool available for an analytical
treatment of this essentially non-Markovian problem.

\begin{figure}[h]
\begin{center}
\def\epsfsize#1#2{0.46\textwidth}
\leavevmode
\epsffile{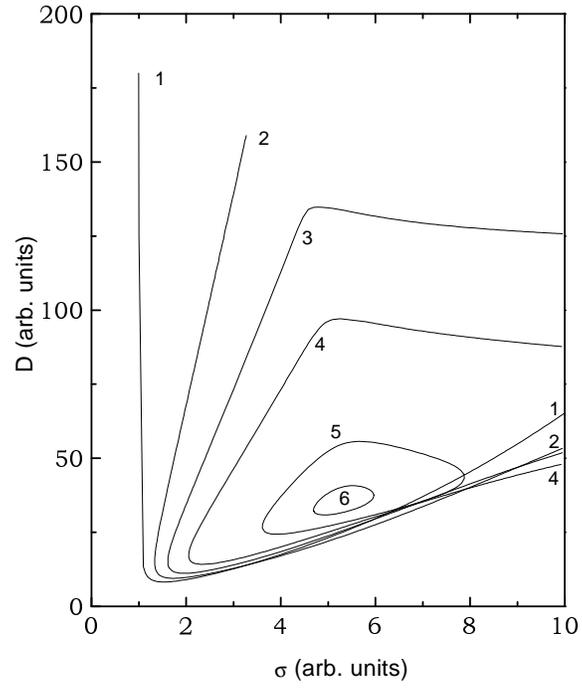}
\end{center}
\caption{\label{mdtw1} Phase diagram in the $\sigma$--$D$ plane, for
different values of $\tau$: (1) $\tau=0$; (2) $\tau=0.015$; (3)
$\tau=0.03$; (4) $\tau=0.05$; (5) $\tau=0.1$; (6) $\tau=0.123$.  The
ordered phase exists only {\em inside\/} the corresponding curves.}
\end{figure}

\begin{figure}[h]
\begin{center}
\def\epsfsize#1#2{0.4\textwidth}
\leavevmode
\epsffile{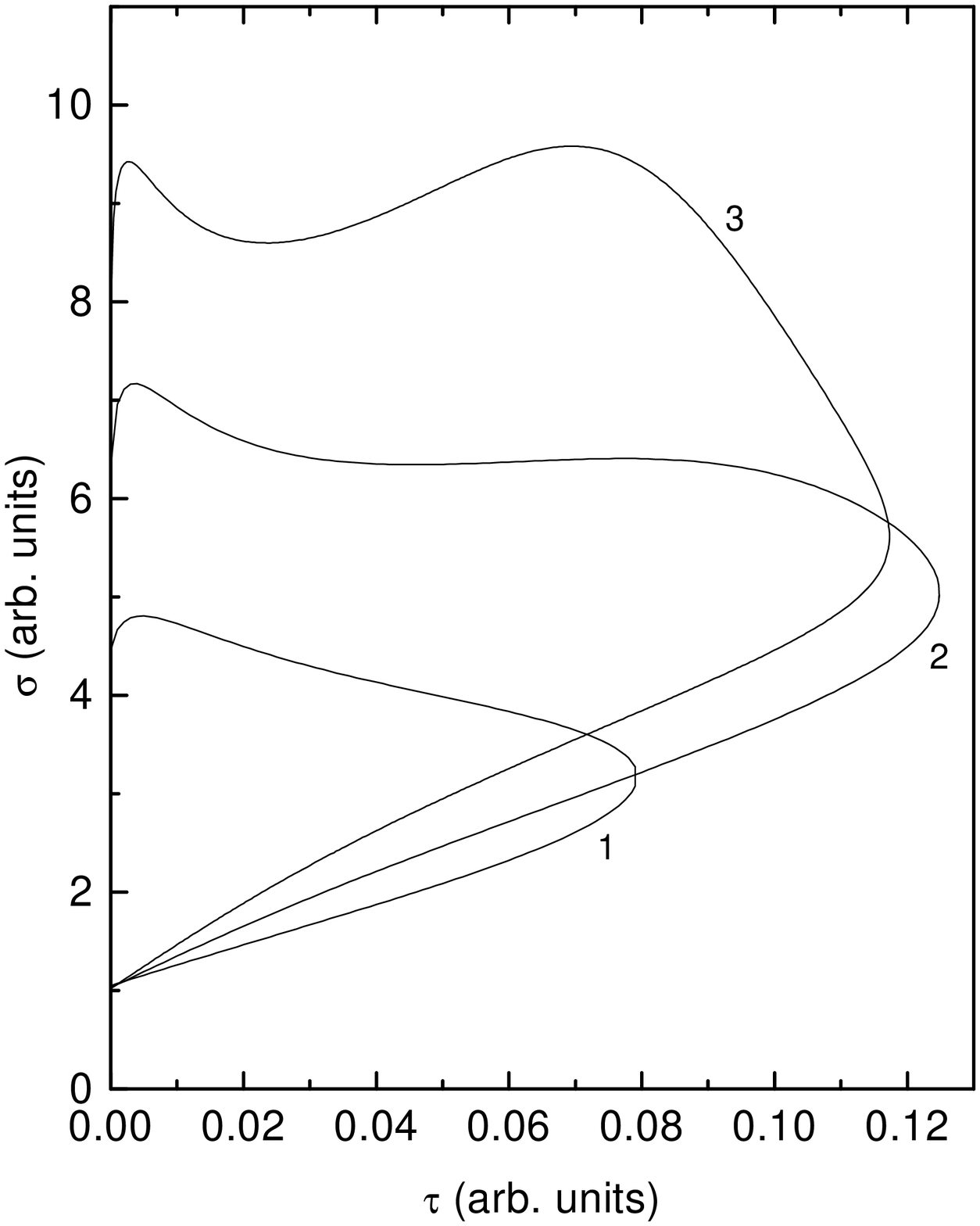}
\end{center}
\caption{\label{mdtw2} Phase diagram in the $\tau$--$\sigma$ plane,
for different values of $D$: (1) $D=20$; (2) $D=32$; (3) $D=45$.  The
ordered phase exists only {\em inside\/} the corresponding curves.}
\end{figure}

\begin{figure}[h]
\begin{center}
\def\epsfsize#1#2{0.4\textwidth}
\leavevmode
\epsffile{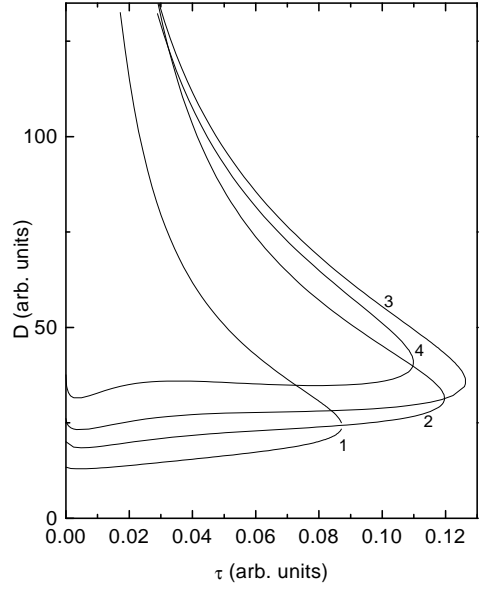}
\end{center}
\caption{\label{mdtw3} Phase diagram in the $\tau$--$D$ plane, for
different values of $\sigma$: (1) $\sigma^2=10$; (2) $\sigma^2=20$;
(3) $\sigma^2=30$; (4) $\sigma^2=50$.  The ordered phase exists only
{\em inside\/} the corresponding curves.}
\end{figure}

\begin{figure}[h]
\begin{center}
\def\epsfsize#1#2{0.4\textwidth}
\leavevmode
\epsffile{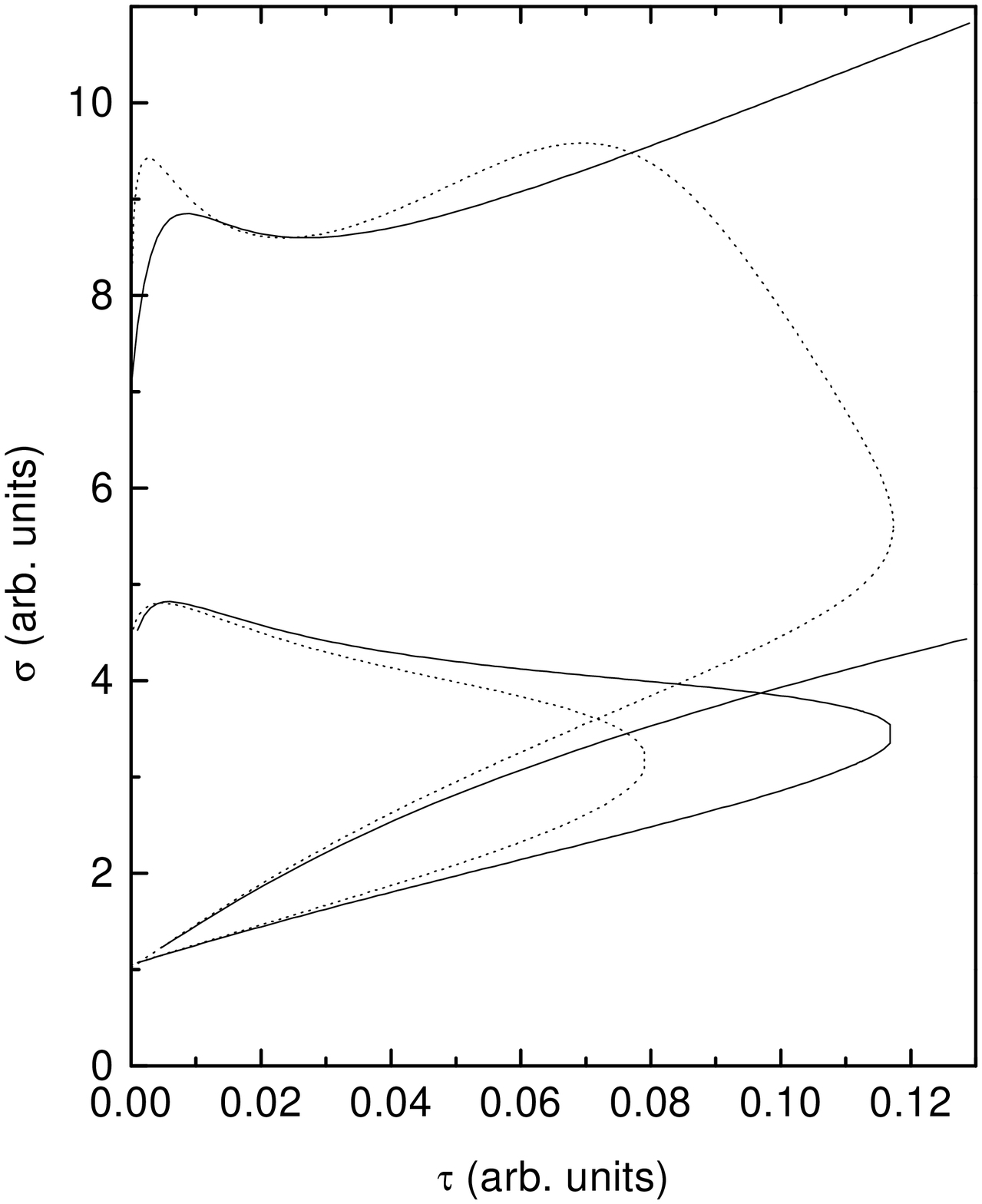}
\end{center}
\caption{\label{mdtw4} Comparison between the simpler mean-field
approach (solid line) and the refined one (dots) in the
$\tau$--$\sigma$ plane, for $D=20$ (lower two curves) and $D=45$
(upper two ones).  Not only does the simpler approach tend to
overestimate the size of the ordered region, but it even predicts
unbounded ordered regions for some values of $D$.}
\end{figure}

\begin{figure}[h]
\begin{center}
\def\epsfsize#1#2{0.4\textwidth}
\leavevmode
\epsffile{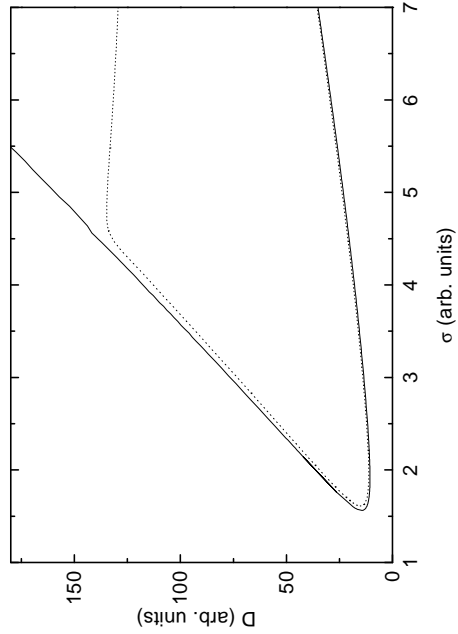}
\end{center}
\caption{\label{mdtw5} Comparison between the two mean-field
approaches in the $\sigma$--$D$ plane, for $\tau=0.03$.  Solid line:
simpler one; dots: refined one.}
\end{figure}

\begin{figure}[h]
\begin{center}
\def\epsfsize#1#2{0.4\textwidth}
\leavevmode
\epsffile{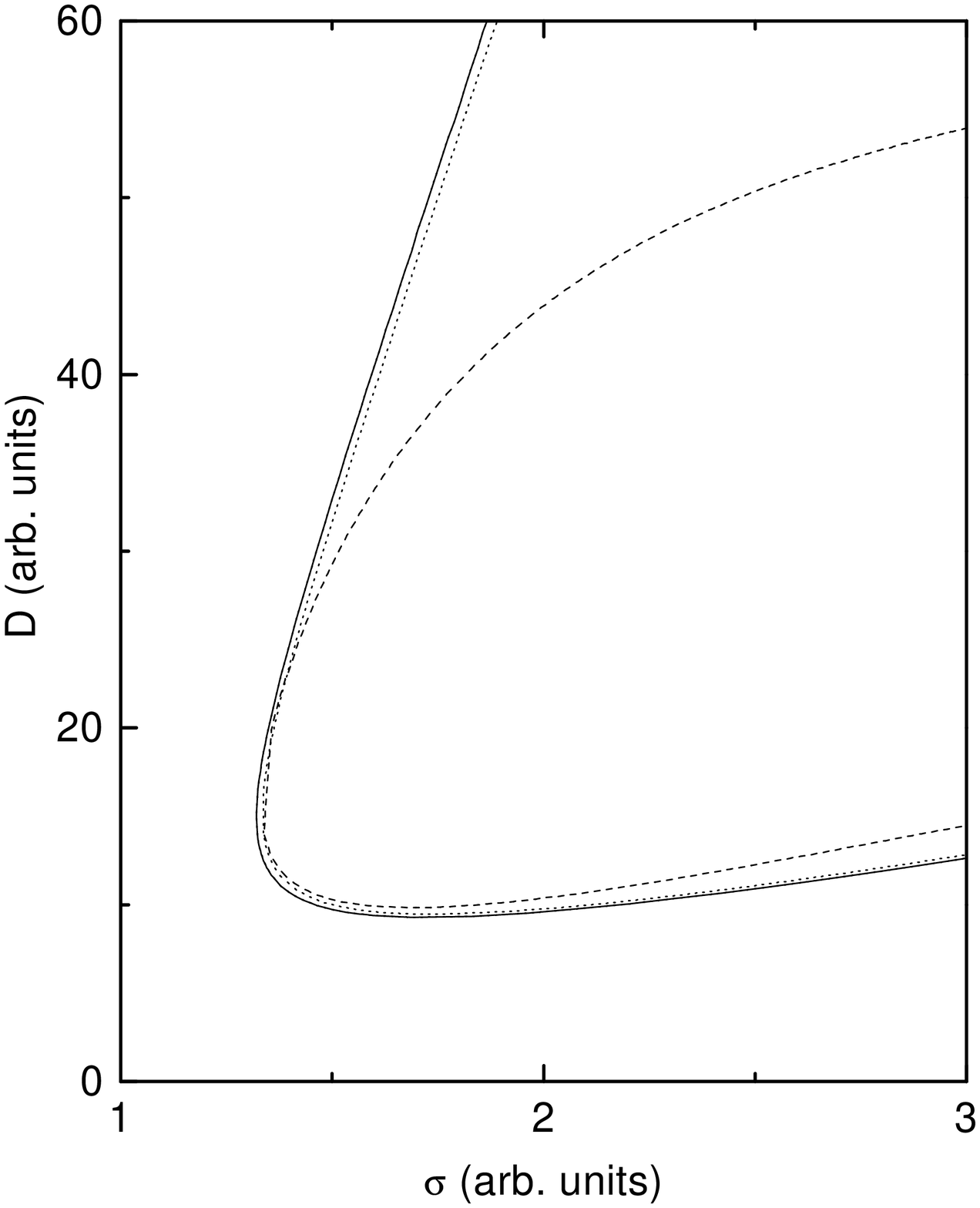}
\end{center}
\caption{\label{mdtw6} Comparison between the two mean-field
approaches and with a perturbative expansion in the $\sigma$--$D$
plane, for $\tau=0.015$.  Solid line: simpler MF; dotted line: refined
MF; dashed line: perturbative.}
\end{figure}

\begin{figure}[h]
\begin{center}
\def\epsfsize#1#2{0.4\textwidth}
\leavevmode
\epsffile{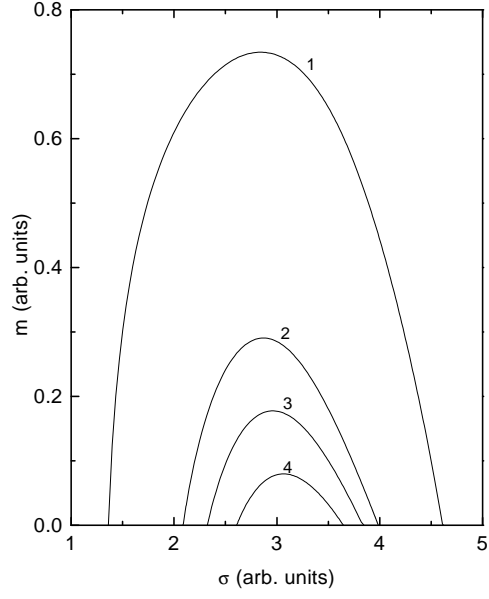}
\end{center}
\caption{\label{mdtw7} Order parameter $m$ vs.\ $\sigma$, for $D=20$
and four values of $\tau$: (1) $\tau=0.015$; (2) $\tau=0.05$; (3)
$\tau=0.06$; (4) $\tau=0.07$.}
\end{figure}

\begin{figure}[h]
\begin{center}
\def\epsfsize#1#2{0.4\textwidth}
\leavevmode
\epsffile{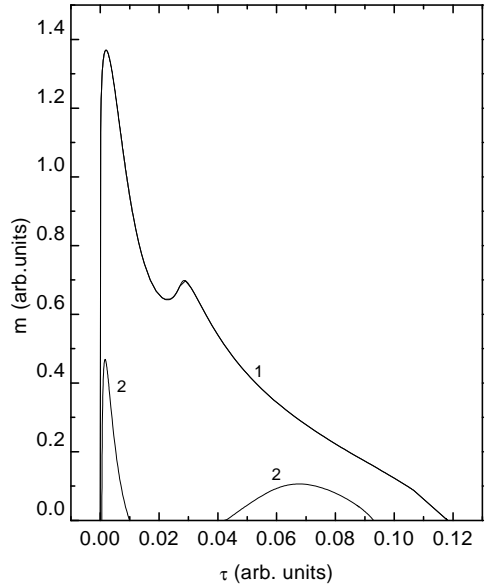}
\end{center}
\caption{\label{mdtw8} Plot of $m$ vs.\ $\tau$ for $D=45$, showing
cases of double reentrance: (1) $\sigma^2=50$; (2) $\sigma^2=80$.}
\end{figure}

\begin{figure}[h]
\begin{center}
\def\epsfsize#1#2{0.4\textwidth}
\leavevmode
\epsffile{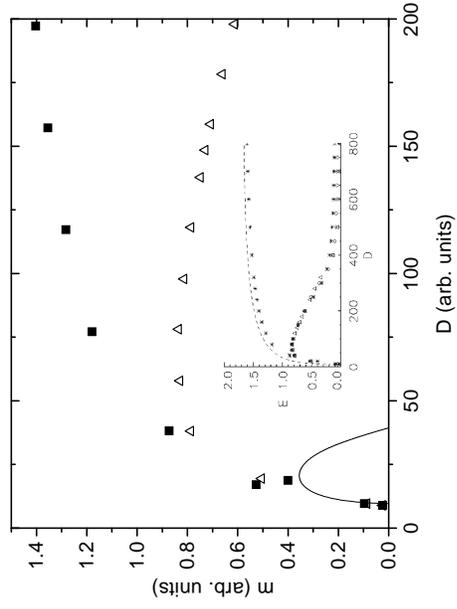}
\end{center}
\caption{\label{mdtw9} Plot of $m$ vs.\ $D$ for $\sigma=2$ and
$\tau=0.01$, showing the predictions of the more refined mean-field
approach together with results coming from a numerical integration of
the original SDE's, Eqs.(\ref{eq:1}).}
\end{figure}

The main result is that for $\tau\neq 0$, the order established as a
consequence of the multiplicative character of the noise can be {\em
destroyed\/} by a strong enough spatial coupling.  Figure \ref{mdtw1}
shows that for given $\tau$ (0.03) and $\sigma$, the ordered phase can
only exist between definite values of $D$.  In particular, the upper
bound on $D$ decreases roughly as $\tau^{-1}$ for given $\sigma$.

The foregoing result can be understood by recalling that the ordered
phase arises as a consequence of the {\em collaboration\/} between the
multiplicative character of the noise and the presence of spatial
coupling.  The disordering effect of $D$ arises {\em only\/} when
$\tau\ne 0$ (the results in Ref.\cite{VPTK}---rightly interpreted in
terms of a ``freezing'' of the short-time behavior by a strong enough
spatial coupling---are thus consistent with ours).  As $\tau$
increases, the minimum value of $D$ required to stabilize the
disordered phase decreases rapidly, and the region in parameter space
available to the ordered phase shrinks until it disappears.

The example worked throughout this paper shows vividly the fact that
the conceptual inheritance from equilibrium thermodynamics (though
often useful) is not always applicable.  The equilibrium-thermodynamic
lore would induce us to think that as $D\to\infty$ an ordered
situation is favored \cite{Pa}.  Although this is certainly true for
the Curie-Weiss model (since in that case the deterministic potential
is itself bistable and an increase of spatial coupling has the effect
of rising the potential barrier between the stable states), it is {\em
not\/} in the case we are dealing with, since the deterministic
potential is {\em monostable}.  Hence, it is the combined effects of
the multiplicative noise {\em and\/} the spatial coupling that induce
the transition.

As a summary, whereas one might say that the value of
Refs.\cite{VPT,VPAH,VPTK} is that they tell experimentalists where
{\em not\/} to look for a noise-induced phase transition---namely, in
those systems which are prone to exhibit a usual (zero-dimensional)
noise-induced transition, and for too large noise intensity---the
present work tells moreover that, due to the consideration of the more
realistic colored noise source, {\em an ordered phase is not to be
found for large values of the spatial coupling either}.  Though the
specific choice of the forms for the functions $f(x)$ and $g(x)$ may
appear to some as physically unmotivated, is up to our knowledge the
{\em simplest\/} one exhibiting this phenomenon.  We believe
nonetheless that the phenomenon is robust and transcends the specific
choice made in this work.

The next obvious step is to consider a finite correlation length in
the lattice model, which requires to go beyond the mean field
approach.  This problem is being presently studied.
\vskip 1.cm
\noindent{\bf Acknowledgments:} R. T. acknowledges financial support
from DGICyT, project numbers PB94-1167 and PB97-0141-C02-01.  H. S. W, R. R.
D. and S. E. M. acknowledge financial support from CONICET, project
number PIP-4953/96, and from ANPCyT, project number .



\end{document}